\documentclass[prl,aps,showpacs,onecolumn]{revtex4}
\usepackage{amsfonts}
\usepackage{graphicx}
\usepackage{amsmath}
\usepackage{bm}

\input{tcilatex}
\begin{document}

\title{Ma-Xu quantization rule and exact WKB condition for translationally
shape invariant potentials}
\author{Y. Grandati and A. B\'{e}rard }
\affiliation{Institut de Physique,, ICPMB, IF CNRS 2843, Universit\'{e} Paul Verlaine, 1
Bd Arago, 57078 Metz, Cedex 3, France}

\begin{abstract}
For translationally shape invariant potentials, the exact quantization rule
proposed by Ma and Xu is a direct consequence of exactness of the modified
WKB quantization condition proved by Barclay. We propose here a very direct
alternative way to calculate the appropriate correction for the whole class
of translationally shape invariant potentials.
\end{abstract}

\maketitle

\section{\protect\bigskip Introduction}

In 2005, Ma and Xu \cite{Maxu1,Maxu2}, on the basis of a previous work of
Cao \cite{Cao,Cao2} and al., proposed a new improved quantization rule
permitting to retrieve the exact spectrum of some exactly solvable quantum
systems. Since, numerous papers have been published on the possible
applications of the Ma-Xu formula to different systems \cite%
{dong,dong2,dong3,Ma,gu,gu2,qiang,qiang2,qiang3,kasri}. Until now, the link
between the exactness of the Ma-Xu formula and the solvability of the
considered system seems to be stayed unclear. In fact, the translationally
shape invariance of the potential suffies to ensure the exactness of the
Ma-Xu formula. This is a direct consequence of a result obtained first by
Barclay\ \cite{barclay2} in 1993 and largely overlooked. He established that
for these specific potentials the higher order terms in the WKB series can
be resummed, yielding to an energy independent correction and he showed that
this Maslov index can be obtained in a closed analytical form. More than ten
years later, Bhaduri and al. \cite{Bhaduri} proposed another interesting
derivation of this result which rests on\ periodic orbit theory (POT) \cite%
{Bhaduri2}.

In the present article, after having established the connection between the
Ma-Xu formula and Barclay's result, we propose an alternative way to
calculate the Maslov index for every translationally shape invariant
potential (TSIP). Starting from a classification of TSIP which uses new
criterions \cite{grandati4}, equivalent to the Barclay-Maxwell ones \cite%
{barclay1}, we show how to obtain this index using simple complex analysis
tools.

\section{Ma-Xu quantization formula}

Consider the stationary one dimensional Schr\"{o}dinger equation ($\hbar
=1,\ m=1/2$):

\begin{equation}
\psi ^{\prime \prime }(x)+p^{2}(x)\psi (x)=0,  \label{eds}
\end{equation}%
where $p(x)=\sqrt{E-V(x)}$ is the classical momentum function for an energy $%
E$. If $x_{1}$ and $x_{2}$ are the classical turning points then $%
p^{2}(x)\geq 0$ for $x\in \left[ x_{1},x_{2}\right] $.

Defining:

\begin{equation}
w(x)=-\frac{\psi ^{\prime }(x)}{\psi (x)},
\end{equation}%
we have:

\begin{equation*}
w^{\prime }(x)=-\frac{\psi ^{\prime \prime }(x)}{\psi (x)}+w^{2}(x).
\end{equation*}

In other words, if $\psi (x)$ satisfies Eq.(\ref{eds}), $w(x)$ is a solution
of the following Riccati equation:

\begin{equation}
w^{\prime }(x)=p^{2}(x)+w^{2}(x)=E-V(x)+w^{2}(x).  \label{edr}
\end{equation}

Every node of $\psi (x)$ (necessarily a simple zero) corresponds to a simple
pole of $w(x)$ which decreases in the interval $\left[ x_{1},x_{2}\right] $.
Consequently, when $x$ runs through $\left[ x_{1},x_{2}\right] $, at each
node of $\psi (x)$ the function $w(x)$ is subject to a discontinuity from $%
-\infty $ to $+\infty $. If we define the phase $\theta (x)$ via:

\begin{equation}
\tan \theta (x)=-\frac{p(x)}{w(x)},  \label{theta}
\end{equation}%
we can write:

\begin{equation}
\theta (x)=\arctan \left( -\frac{p(x)}{w(x)}\right) +n\pi ,  \label{theta3}
\end{equation}%
where $\arctan y\in \left[ -\pi /2,\pi /2\right] $ is the principal
determination of the reciprocal of the tangent function and where $n$
increases of 1 at each node of $w(x)$.

Then, if $x_{1}$ and $x_{2}$ are not nodes of $\psi (x)$:

\begin{equation}
\int_{x_{1}}^{x_{2}}\theta ^{\prime }(x)dx=N\pi ,
\end{equation}%
$N$ being the total number of nodes of $w(x)$ on $\left[ x_{1},x_{2}\right] $%
.

But using Eq.(\ref{edr}) we also have:

\begin{equation}
\theta ^{\prime }(x)=p(x)-\frac{p^{\prime }(x)w(x)}{w^{\prime }(x)}.
\end{equation}

Considering the n$^{th}$ bound state at energy $E_{n}$ for which $N_{n}=n+1$%
, we obtain the following identity:

\begin{equation}
\int_{x_{1,n}}^{x_{2,n}}p_{n}(x)dx=n\pi +\gamma (E_{n}),  \label{exactwkb}
\end{equation}%
where the correction $\gamma (E_{n})$ to the lowest order WKB\ condition is
given by the following integral:

\begin{equation}
\gamma (E_{n})=\pi +\int_{x_{1,n}}^{x_{2,n}}\frac{w_{n}(x)p_{n}^{\prime }(x)%
}{w_{n}^{\prime }(x)}dx.
\end{equation}

Ma and Xu \cite{Maxu1,Maxu2} have observed that que for a large class of
exactly solvable potentials, this integral correction is in fact independent
of the considered energy level, and can be calculated from the ground state:

\begin{equation}
\gamma (E_{n})=\gamma (E_{0})=\pi +\int_{x_{1,0}}^{x_{2,0}}\frac{%
w_{0}(x)p_{0}^{\prime }(x)}{w_{0}^{\prime }(x)}dx.  \label{correc}
\end{equation}

Inserting Eq.(\ref{correc}) in Eq.(\ref{exactwkb}) gives the Ma-Xu formula:

\begin{equation}
\int_{x_{1,n}}^{x_{2,n}}p_{n}(x)dx=n\pi +\gamma (E_{0}).  \label{maxu}
\end{equation}

The knowledge of the ground state characteristics permits then to calculate
the $\gamma (E_{0})$ correction and reaches to an implicit formula for the
other energy levels $E_{n}$. The explicit calculations have been performed
for many exactly solvable examples \cite%
{dong,dong2,dong3,Ma,gu,gu2,qiang,qiang2,qiang3,kasri}. Nevertheless the
question of a direct link between the exact solvability of the potential and
the validity of the above formula is still stayed open.

As much as we know, all the closed form analytically solvable quantum
mechanical systems\ belongs to the set of translationally shape invariant
potentials (TSIP) in the sense of SUSY quantum mechanics \cite{cooper}. In
fact for this set, the validity of the Ma-Xu formula follows from a result
established in 1993 by Barclay \cite{barclay2}. In this article, he
established that for these potentials, the WKB series can be resummed beyond
the lowest order giving an energy independent correction\ which can be
absorbed in the Maslov index and written in a closed analytical form. He
showed equally that this result is directly correlated to the exactness of
the lowest order SWKB quantization condition \cite{cooper,comtet}. The
starting point is the definition of two classes of potentials, each
characterized by a specific change of variable which brings the potential
into a quadratic form. This two classes are shown to coincide to the
Barclay-Maxwell classes \cite{barclay1} which are based on a functional
characterization of superpotentials and which cover the whole set of TSIP.
In both cases, the action variable for an energy $E_{n}$ is of the form:

\begin{equation}
I_{n}=\doint\nolimits_{E_{n}}p_{n}(x)dx=2%
\int_{x_{1,n}}^{x_{2,n}}p_{n}(x)dx=2\left( n\pi +\gamma \right) ,
\end{equation}%
where $\gamma $ is an energy-independent correction characteristic of the
considered potential. As shown later by Bhaduri and al. \cite{Bhaduri}, this
results can be retrieved in an elegant way using POT \cite{Bhaduri2}.

The translational shape invariance of the potential is then a sufficient
condition for the validity of the Ma-Xu prescription which reaches to Eq.(%
\ref{maxu}).

In the following, we propose an alternative way to recover Barclay's result.
It lies on a different characterization of the Barclay-Maxwell classes that
we recently presented \cite{grandati4}. In both cases the action variable
can be rewritten as a complex integral on an uniformization domain including
only one branch-cut. Using standard tools of complex analysis, it can be
readily calculated to recover the explicit form of $\gamma $ for the whole
set of TSIP.

Note that analogous complex analysis techniques have been employed by Bhalla
and al \cite{bhalla,bhalla2,bhalla3} in the frame of the quantum
Hamilton-Jacobi (QHJ) formalism \cite{leacock1,leacock2} where, starting
from the exact QHJ quantization condition, they have computed the spectrum
of numerous potentials for which they have also showed the exactness of the
lowest order SWKB quantization condition. Moreover, Cherqui et al \cite%
{cherqui} have recently shown that for algebraic and hyperbolic
translationally shape invariant potentials, the shape invariance condition
provides sufficient information on the singularity structure of the quantum
momentum function to determine directly the energy spectrum of the system
from the exact QHJ quantization condition.

\section{First category potentials}

We say that a one dimensional potential is of first category \cite{grandati4}
if there exists a change of variable $x\rightarrow u$ transforming the
potential into an harmonic one $V(x)\rightarrow V(u)=\widetilde{\lambda }%
_{2}u^{2}+\widetilde{\lambda }_{1}u+\widetilde{\lambda }_{0}$, such that $%
u(x)$ satisfy a constant coefficient Riccati equation:

\begin{equation}
\frac{du(x)}{dx}=A_{0}+A_{1}u(x)+A_{2}u^{2}(x),  \label{potreg}
\end{equation}%
$du/dx$ being of constant sign in all the range of values of $x$ and $u$.

The one dimensional harmonic oscillator correspond to the special case $%
A_{1}=A_{2}=0$ and the Morse potential is generically associated to the case 
$A_{1}\neq 0,\ A_{2}=0$.

\subsection{Harmonic oscillator}

It is perfectly well known that the EBK quantization condition is exact for
the harmonic oscillator implying then the exactness of the Ma-Xu formula
with a Maslov index equal to $1/2$. Nevertheless, we will examine this case
completely as a first example of use of the complex variable integration
technique.

The harmonic oscillator potential with zero ground state energy is:

\begin{equation}
V(x,\omega )=\left( \omega /2\right) ^{2}x^{2}-\omega /2.
\end{equation}

The classical turning points $\pm x_{0,n}$ at energy $E_{n}=n\omega $ 
satisfy:

\begin{equation}
x_{0,n}^{2}=\frac{4}{\omega }\left( n+\frac{1}{2}\right) .
\end{equation}

The half action variable for a classical periodic orbit of energy $E_{n}$ is
then:

\begin{equation}
I_{n}=\int_{x_{1,n}}^{x_{2,n}}p_{n}(x)dx=\frac{\omega }{2}%
\int_{-x_{0,n}}^{x_{0,n}}dx\sqrt{x_{0,n}^{2}-x^{2}}.
\end{equation}

The complex function $f_{H}(z)=\sqrt{x_{0,n}^{2}-z^{2}}$, defined on $%
\mathbb{C-}\left] -x_{0,n},x_{0,n}\right[ $, admits only one isolated
singularity at infinity. Then:

\begin{equation}
\int_{-x_{0,n}}^{x_{0,n}}dxf_{H}(x)=\pi i\text{Res}\left( f_{H}(z),\infty
\right) .
\end{equation}

The asymptotic behaviour at infinity on the Riemann sheet for which $\sqrt{%
u_{0,n}^{2}-z^{2}}=\sqrt{x_{0,n}^{2}+y^{2}}>0$ on the positive half
imaginary axe ($z=iy$, $y>0$) is given by:

\begin{equation}
\sqrt{x_{0,n}^{2}-z^{2}}\underset{z\rightarrow \infty }{\sim }-iz\left( 1-%
\frac{x_{0,n}^{2}}{z^{2}}\right) ^{\frac{1}{2}}\underset{z\rightarrow \infty 
}{\sim }-iz+\frac{ix_{0,n}^{2}}{2}\frac{1}{z}+O(\frac{1}{z^{2}}).
\end{equation}

Consequently:

\begin{equation}
\text{Res}\left( f_{H}(z),\infty \right) =-\frac{ix_{0,n}^{2}}{2}
\end{equation}%
and:

\begin{equation}
\int_{x_{1,n}}^{x_{2,n}}p_{n}(x)dx=\left( n+\frac{1}{2}\right) \pi .
\end{equation}%
which is the expected result.

\subsection{Morse potential}

As for the harmonic oscillator, the EBK formula is exact for the Morse
potential. This second example is nevertheless very instructive. The Morse
potential with zero energy ground state is \cite{cooper}:

\begin{equation}
V(x)=A^{2}+B^{2}e^{-2\alpha x}-2B\left( A+\frac{\alpha }{2}\right)
e^{-\alpha x},\ \alpha >0.  \label{potMorse}
\end{equation}

Using the change of variable $y=\exp \left( -\alpha x\right) ,\ dy=-\alpha
ydx$, it becomes:

\begin{equation}
V(y)=B^{2}\left( y-y_{0}\right) ^{2}+V_{0}
\end{equation}%
with $\ y_{0}=\left( A+\alpha /2\right) /B$ and:

\begin{equation}
V_{0}=A^{2}-B^{2}y_{0}^{2}=\left( A+\frac{\alpha }{2}\right) ^{2}-A^{2}.
\end{equation}

This is a translationally shape invariant potential with a spectrum given by 
\cite{grandati4}:

\begin{equation}
\ E_{n}=a^{2}-a_{n}^{2}.  \label{SIPMorse}
\end{equation}%
with $a=A$ and $a_{k}=A-k\alpha $.

As for the classical turning points, they are given by:

\begin{equation}
\left\{ 
\begin{array}{c}
y_{2,n}=y_{0}+u_{0,n} \\ 
y_{1,n}=y_{0}-u_{0,n}%
\end{array}%
\right.
\end{equation}%
with:

\begin{equation}
u_{0,n}=\sqrt{y_{0}^{2}-\frac{a_{n}^{2}}{B^{2}}}.
\end{equation}

The half action variable for a classical periodic orbit of energy $E_{n}$ is
then:

\begin{eqnarray}
\int_{x_{1,n}}^{x_{2,n}}p_{n}(x)dx &=&-\frac{1}{\alpha }%
\int_{y_{1,n}}^{y_{2,n}}\sqrt{E_{n}-V(y,a)}\frac{dy}{y} \\
&=&\frac{B}{\alpha }\int_{-u_{0,n}}^{u_{0,n}}f_{M}\left( u\right) du,  \notag
\end{eqnarray}%
where:

\begin{equation}
f_{M}\left( u\right) =\frac{\sqrt{u_{0,n}^{2}-u^{2}}}{u+y_{0}}.
\end{equation}

The complex extended function $f_{M}\left( z\right) =\sqrt{u_{0,n}^{2}-z^{2}}%
/\left( z+y_{0}\right) T,$ defined on $\mathbb{C-}\left] -u_{0,n},u_{0,n}%
\right[ $, admits two isolated singularities at infinity and in $-y_{0}$.
Then:

\begin{equation}
\int_{-u_{0,n}}^{u_{0,n}}duf_{M}\left( u\right) =\pi i\left( \text{Res}%
\left( f_{M}\left( z\right) ,-y_{0}\right) +\text{Res}\left( f_{M}\left(
z\right) ,\infty \right) \right) .
\end{equation}

With the same choice as before for the square root determination we have:

\begin{equation}
f_{M}\left( z\right) \underset{z\rightarrow \infty }{\sim }i\left( -1+\frac{%
y_{0}}{z}+O(\frac{1}{z^{2}})\right)
\end{equation}%
which implies:

\begin{equation}
\text{Res}\left( f_{M}\left( z\right) ,\infty \right) =-i\frac{a+\alpha /2}{B%
}.
\end{equation}

The residue in $-y_{0}$ (which is a simple pole) is readily obtained as:%
\begin{equation}
\text{Res}\left( f_{M}\left( z\right) ,-y_{0}\right) =i\frac{a-n\alpha }{B}.
\end{equation}

Then:

\begin{equation}
\int_{x_{1,n}}^{x_{2,n}}p_{n}(x)dx=\left( n+\frac{1}{2}\right) \pi .
\end{equation}%
which is the expected result.

\subsection{Effective radial potential for the Kepler-Coulomb problem}

Consider the effective radial potential for the Kepler-Coulomb. It can be
written as \cite{grandati4}:

\begin{equation}
V(y)=l(l+1)y^{2}-\gamma y+\frac{\gamma ^{2}}{4(l+1)^{2}},\ k>0,
\end{equation}%
where $y=1/x$ and$\ dy=-y^{2}dx$. We can rather use the equivalent form:

\begin{equation}
V(y)=l(l+1)\left( y-y_{0}\right) ^{2}+V_{0},\ 
\end{equation}%
with $y_{0}=\gamma /2l(l+1)$ and 
\begin{equation}
V_{0}=\frac{\gamma ^{2}}{4(l+1)^{2}}-\frac{\gamma ^{2}}{4l(l+1)}.
\end{equation}%
.

This is a  translationally shape invariant potential characterized by the
following energy spectrum \cite{grandati4}:

\begin{equation}
E_{n}(a)=\frac{\gamma ^{2}}{4a^{2}}-\frac{\gamma ^{2}}{4a_{n}^{2}}.
\end{equation}%
where $a=l+1$ and $a_{k}=l+1+k,$.

The classical turning points are given by:

\begin{equation}
\left\{ 
\begin{array}{c}
y_{2,n}=y_{0}+u_{0,n} \\ 
y_{1,n}=y_{0}-u_{0,n}%
\end{array}%
\right.
\end{equation}%
with:

\begin{equation}
u_{0,n}=\frac{1}{\sqrt{l(l+1)}}\sqrt{\frac{\gamma ^{2}}{4l(l+1)}-\frac{%
\gamma ^{2}}{4a_{n}^{2}}}.
\end{equation}

The half action variable for a classical periodic orbit of energy $E_{n}$ is
then:

\begin{eqnarray}
\int_{x_{1,n}}^{x_{2,n}}p_{n}(x)dx &=&-\int_{y_{1,n}}^{y_{2,n}}\sqrt{%
E_{n}-V(y,a)}\frac{dy}{y^{2}} \\
&=&-\sqrt{l(l+1)}\int_{-u_{0,n}}^{u_{0,n}}duf_{K}\left( u\right) ,  \notag
\end{eqnarray}%
where:

\begin{equation}
f_{K}\left( u\right) =\frac{\sqrt{u_{0,n}^{2}-u^{2}}}{\left( u+y_{0}\right)
^{2}}.
\end{equation}

Again, the complex function $f_{K}\left( z\right) =\sqrt{u_{0,n}^{2}-z^{2}}%
/\left( z+y_{0}\right) ^{2}$, defined on $\mathbb{C-}\left] -u_{0,n},u_{0,n}%
\right[ $, admits two isolated singularities at infinity and in $-y_{0}$.
Then:

\begin{equation}
\int_{-u_{0,n}}^{u_{0,n}}duf_{K}\left( u\right) =\pi i\left( \text{Res}%
\left( f_{K}\left( z\right) ,-y_{0}\right) +\text{Res}\left( f_{K}\left(
z\right) ,\infty \right) \right) .
\end{equation}

With the same choice as before for the square root determination, we have:

\begin{equation}
f_{K}\left( z\right) \underset{z\rightarrow \infty }{\sim }\frac{-i}{z}%
\left( 1-\frac{u_{0,n}^{2}}{z^{2}}\right) ^{\frac{1}{2}}\left( 1+\frac{y_{0}%
}{z}\right) ^{-2}\underset{z\rightarrow \infty }{\sim }\frac{-i}{z}+O(\frac{1%
}{z^{2}})
\end{equation}%
which implies:

\begin{equation}
\text{Res}\left( f_{K}\left( z\right) ,\infty \right) =i.
\end{equation}

The residue in $-y_{0}$ which is a double pole is readily obtained as:%
\begin{equation}
\text{Res}\left( f_{K}\left( z\right) ,-y_{0}\right) =i\frac{a_{n}}{\sqrt{%
l(l+1)}}.
\end{equation}

Then:

\begin{equation}
\int_{x_{1,n}}^{x_{2,n}}p_{n}(x)dx=n\pi +\gamma ,
\end{equation}%
where:

\begin{equation*}
\gamma =\pi \left( l+1+\sqrt{l(l+1)}\right)
\end{equation*}%
which is the expected result.

\subsection{Other first category potentials}

If we except the two preceding examples, all the other first category
potentials are such that there exists a change of variable $x\rightarrow y$
transforming the potential into an harmonic one:

\begin{equation*}
V(x)\rightarrow V(y)=a\left( a\mp \alpha \right) y^{2}+\lambda _{1}y+\lambda
_{0}(a),
\end{equation*}%
with $\lambda _{0}\left( a\right) =\lambda _{1}^{2}/4a^{2}-\alpha a$ (in
order to have a zero energy ground state) and where $y(x)$ satisfies a
constant coefficient Riccati equation of the form \cite{grandati4}:

\begin{equation}
\frac{dy}{dx}=\alpha \pm \alpha y^{2}(x)>0.  \label{potreg2}
\end{equation}

\bigskip The potential can still be written \cite{grandati4}:

\begin{equation}
V_{\pm }(y,a)=a\left( a\mp \alpha \right) \left( y-y_{0}\right) ^{2}+V_{0},
\label{potcat1}
\end{equation}%
where:

\begin{equation}
y_{0}=-\lambda _{1}/a\left( a\mp \alpha \right) ,\quad V_{0}=\lambda
_{0}(a)-a\left( a\mp \alpha \right) y_{0}^{2}
\end{equation}%
.

$V$ is a  translationally shape invariant potential and its energy spectrum
is then given by \cite{grandati4}:

\begin{equation}
\ E_{n}(a)=\phi _{1,\pm }\left( a\right) -\phi _{1,\pm }\left( a_{n}\right) ,
\label{spectrecat1}
\end{equation}%
where $a_{k}=a\pm k\alpha $ and:

\begin{equation}
\phi _{1,\pm }\left( a\right) =\mp a^{2}+\frac{\lambda _{1}^{2}}{4a^{2}}.
\label{phi1}
\end{equation}

The classical turning points $y_{i,n}$ for an energy $E_{n}$ are determined
by the condition:

\begin{equation}
E_{n}=V(y_{i,n})\Leftrightarrow E_{n}=a\left( a\mp \alpha \right)
u_{0,n}^{2}+V_{0},  \label{turning points}
\end{equation}%
where:

\begin{equation}
\left\{ 
\begin{array}{c}
y_{2,n}=y_{0}+u_{0,n} \\ 
y_{1,n}=y_{0}-u_{0,n}%
\end{array}%
\right. ,  \label{centr1}
\end{equation}%
and:

\begin{equation}
u_{0,n}=\sqrt{\mp 1+y_{0}^{2}-\frac{\phi _{1,\pm }\left( a_{n}\right) }{%
a\left( a\mp \alpha \right) }.}  \label{centr2}
\end{equation}

Consider first the case where $dy/dx=\alpha +\alpha y^{2}(x)$. The half
action variable for a classical periodic orbit of energy $E_{n}$ is given by:

\begin{eqnarray}
\int_{x_{1,n}}^{x_{2,n}}p_{n}(x)dx &=&\int_{y_{1,n}}^{y_{2,n}}\sqrt{%
E_{n}-V(y,a)}\frac{dy}{\alpha \left( y-i\right) \left( y+i\right) } \\
&=&\frac{\sqrt{a\left( a-\alpha \right) }}{\alpha }%
\int_{-u_{0,n}}^{u_{0,n}}f_{1,+}(u)du,  \notag
\end{eqnarray}%
where:

\begin{equation}
f_{1,+}(z)=\frac{\sqrt{u_{0,n}^{2}-z^{2}}}{\left( z-z_{0}\right) \left( z-%
\overline{z_{0}}\right) }
\end{equation}%
with $z_{0}=-y_{0}+i$.

The complex extended integrand $f_{1,+}(z)$, defined on $\mathbb{C-}\left]
-u_{0,n},u_{0,n}\right[ $, admits three isolated singularities at infinity, $%
z_{0}$ and $\overline{z_{0}}$. Then:

\begin{equation}
\int_{-u_{0,n}}^{u_{0,n}}duf_{1,+}(u)=\pi i\left( \text{Res}\left(
f_{1,+}(z),z_{0}\right) +\text{Res}\left( f_{1,+}(z),\overline{z_{0}}\right)
+\text{Res}\left( f_{1,+}(z),\infty \right) \right) .
\end{equation}

When $z\rightarrow \infty $:

\begin{equation}
f_{1,+}(z)\underset{z\rightarrow \infty }{\sim }-\frac{i}{z}+O(\frac{1}{z^{2}%
}),
\end{equation}%
which implies:

\begin{equation}
\text{Res}\left( f_{1,+}(z),\infty \right) =i.
\end{equation}

As for the residues in $z_{0}$ and $\overline{z_{0}}$, using Eq.(\ref%
{turning points}), Eq.(\ref{centr1}), Eq.(\ref{centr2}) and Eq.(\ref%
{spectrecat1}) we deduce:

\begin{equation}
\text{Res}\left( f_{1,+}(z),z_{0}\right) =-\frac{1}{2\sqrt{a\left( a-\alpha
\right) }}\left( ia_{n}+\frac{\lambda _{1}}{2a_{n}}\right)
\end{equation}%
and:

\begin{equation}
\text{Res}\left( f_{1,+}(z),\overline{z}_{0}\right) =-\frac{1}{2\sqrt{%
a\left( a-\alpha \right) }}\left( -ia_{n}+\frac{\lambda _{1}}{2a_{n}}\right)
.
\end{equation}

Then:

\begin{equation}
\int_{x_{1,n}}^{x_{2,n}}p_{n}(x)dx=n\pi +\gamma
\end{equation}%
with:

\begin{equation}
\gamma =\frac{\pi a}{\alpha }\left( 1-\sqrt{1-\frac{\alpha }{a}}\right) .
\end{equation}

Consider now the case $dy/dx=\alpha -\alpha y^{2}(x)$. The half action
variable for a classical periodic orbit of energy $E_{n}$ is:

\begin{eqnarray}
\int_{x_{1,n}}^{x_{2,n}}p_{n}(x)dx &=&\int_{y_{1,n}}^{y_{2,n}}\sqrt{%
E_{n}-V(y,a)}\frac{dy}{\alpha \left( 1-y\right) \left( 1+y\right) } \\
&=&-\frac{\sqrt{a\left( a-\alpha \right) }}{\alpha }%
\int_{-u_{0,n}}^{u_{0,n}}f_{1,-}(u)du,  \notag
\end{eqnarray}%
where:

\begin{equation}
f_{1,-}(u)=\frac{\sqrt{u_{0,n}^{2}-u^{2}}}{\left( u-z_{1}\right) \left(
u-z_{2}\right) }
\end{equation}%
with $z_{1}=-1-y_{0}$ and $z_{2}=1-y_{0}$.

The complex extended integrand $f_{1,+}(z)$, defined on $\mathbb{C-}\left]
-u_{0,n},u_{0,n}\right[ $, admits three isolated singularities at infinity, $%
z_{1}$ and $z_{2}$. Then:

\begin{equation}
\int_{-u_{0,n}}^{u_{0,n}}duf_{1,-}(u)=\pi i\left( \text{Res}\left(
f_{1,-}(z),z_{1}\right) +\text{Res}\left( f_{1,-}(z),z_{2}\right) +\text{Res}%
\left( f_{1,-}(z),\infty \right) \right) .
\end{equation}

When $z\rightarrow \infty $:

\begin{equation}
f_{1,+}(z)\underset{z\rightarrow \infty }{\sim }-\frac{i}{z}+O(\frac{1}{z^{2}%
}),
\end{equation}%
which implies:

\begin{equation}
\text{Res}\left( f_{1,+}(z),\infty \right) =i.
\end{equation}

As for the residues in $z_{0}$ and $\overline{z_{0}}$, from Eq.(\ref{turning
points}), Eq.(\ref{centrage1}), Eq.(\ref{centrage2}) and Eq.(\ref%
{spectrecat1}) we deduce:

\begin{equation}
\text{Res}\left( f_{1,-}(z),z_{2}\right) =\frac{i}{2\sqrt{a\left( a-\alpha
\right) }}\left( a_{n}+\frac{\lambda _{1}}{2a_{n}}\right)
\end{equation}%
and:

\begin{equation}
\text{Res}\left( f_{1,-}(z),z_{1}\right) =\frac{i}{2\sqrt{a\left( a-\alpha
\right) }}\left( a_{n}-\frac{\lambda _{1}}{2a_{n}}\right) .
\end{equation}

Note that in this case, we had to change the square root determination when
we pass from $z_{1}$ to $z_{2}$, since we have either $-1<y_{1,n}<y_{2,n}$ $%
<1$(when $y=\tanh (\alpha x+\varphi )$), that is $%
-1-y_{0}<-u_{0,n}<u_{0,n}<1-y_{0}$, or $y_{1,n}<-1$ $<1<y_{2,n}$ (when $%
y=\coth (\alpha x+\varphi )$), that is $-u_{0,n}<-1-y_{0}<1-y_{0}<u_{0,n}$.

Then:

\begin{equation}
\int_{x_{1,n}}^{x_{2,n}}p_{n}(x)dx=n\pi +\gamma
\end{equation}%
with:

\begin{equation}
\gamma =-\frac{\pi a}{\alpha }\left( 1+\sqrt{1-\frac{\alpha }{a}}\right) .
\end{equation}

We recover the results obtained in \cite{Bhaduri} for the first class
potentials.

\section{Second category potentials}

We say that a one dimensional potential is of second category if there
exists a change of variable $x\rightarrow u$ transforming the potential into
an isotonic one $V(x)\rightarrow V(u)=\widetilde{\lambda }_{2}u^{2}+%
\widetilde{\lambda }_{0}+\frac{\widetilde{\mu }_{2}}{u^{2}}$, such that $u(x)
$ satisfies a constant coefficients Riccati equation of the form:

\begin{equation}
\frac{du(x)}{dx}=A_{0}+A_{2}u^{2}(x),  \label{potsing}
\end{equation}%
$du(x)/dx$ being of constant sign in all the range of values of $x$ and $u$.

\subsection{Isotonic oscillator}

The isotonic potential with a zero energy ground state $\left(
E_{0}=0\right) $ is \cite{weissman,grandati4}:

\begin{equation}
V_{-}(x)=\frac{\omega ^{2}}{4}x^{2}+\frac{l(l+1)}{x^{2}}-\omega \left( l+%
\frac{3}{2}\right) ,\ l>0.
\end{equation}

Its energy spectrum is given by \cite{grandati4}:

\begin{equation}
E_{n}(a)=2n\omega ,
\end{equation}%
where $a=\left( \frac{\omega }{2},l+1\right) $ and $a_{k}=\left( \frac{%
\omega }{2},l+1+k\right) $.

The classical turning points $x_{i,n}$ for an energy $E_{n}$ are determined
by the condition:

\begin{equation}
E_{n}=V(x_{i,n})\Leftrightarrow \frac{\omega ^{2}}{4}x_{i,n}^{4}-\omega
\left( 2n+l+\frac{3}{2}\right) x_{i,n}^{2}+l(l+1)=0,
\end{equation}%
that is:

\begin{equation}
\left\{ 
\begin{array}{c}
x_{2,n}^{2}=u_{0,n}+\delta _{n} \\ 
x_{1,n}^{2}=u_{0,n}-\delta _{n}%
\end{array}%
\right.
\end{equation}%
with:

\begin{equation}
u_{0,n}=2\frac{2n+l+\frac{3}{2}}{\omega },\ \delta _{n}=\frac{2}{\omega }%
\sqrt{\left( 2n+l+\frac{3}{2}\right) ^{2}-l(l+1)}.
\end{equation}

The half action variable for a classical periodic orbit of energy $E_{n}$ is
then:

\begin{equation}
\int_{x_{1,n}}^{x_{2,n}}p_{n}(x)dx=\frac{\omega }{4}%
\int_{x_{1,n}^{2}}^{x_{2,n}^{2}}duf_{I}(u),  \notag
\end{equation}%
where we have defined $u=x^{2}$ and:

\begin{equation*}
f_{I}(u)=\frac{\sqrt{\left( u-x_{1,n}^{2}\right) \left( x_{2,n}^{2}-u\right) 
}}{u}.
\end{equation*}

The complex extended function $f_{I}(z)=\sqrt{\left( z-x_{1,n}^{2}\right)
\left( x_{2,n}^{2}-z\right) }/z$, defined on $\mathbb{C-}\left]
x_{1,n}^{2},x_{2,n}^{2}\right[ $, admits two isolated singularities at
infinity and in $0$. Then:

\begin{equation}
\int_{x_{1,n}^{2}}^{x_{2,n}^{2}}duf_{I}(u)=\pi i\left( \text{Res}\left(
f_{I}(z),0\right) +\text{Res}\left( f_{I}(z),\infty \right) \right) .
\end{equation}

When $z\rightarrow \infty $:

\begin{equation}
f_{I}(z)\underset{z\rightarrow \infty }{\sim }-i\left( 1-\frac{%
x_{2,n}^{2}+x_{1,n}^{2}}{2z}+O(\frac{1}{z^{2}})\right) ,
\end{equation}%
which implies:

\begin{equation}
\text{Res}\left( f_{I}(z),\infty \right) =-iu_{0,n}=-2i\frac{2n+l+\frac{3}{2}%
}{\omega ^{2}}.
\end{equation}

As for the residue in $0$, from Eq.(\ref{turning points}) we have:%
\begin{equation}
\text{Res}\left( f_{I}(z),0\right) =\sqrt{-x_{1,n}^{2}x_{2,n}^{2}}=i\frac{2}{%
\omega }\sqrt{l(l+1)}.
\end{equation}

Then:

\begin{equation}
\int_{x_{1,n}}^{x_{2,n}}p_{n}(x)dx=n\pi +\gamma ,
\end{equation}%
with:

\begin{equation}
\gamma =\frac{\pi }{2}\left( l+\frac{3}{2}-\sqrt{l(l+1)}\right) .
\label{maslovIsot}
\end{equation}

\subsection{\protect\bigskip Other second category potentials}

If we except the preceding example, all the other second category potentials
are such that there exists a change of variable $x\rightarrow y$
transforming the potential into an isotonic one $V(x)\rightarrow
V(y)=\lambda _{2}y^{2}+\frac{\mu _{2}}{y^{2}}+\lambda _{0}$, where $y(x)>0$
satisfies a constant coefficient Riccati equation of the form \cite%
{grandati4}:

\begin{equation}
\frac{dy}{dx}=\alpha \pm \alpha y^{2}(x),  \label{potsing2}
\end{equation}%
$dy(x)/dx$ being of constant sign in all the range of values of $x$ and $y$.

\bigskip As shown in \cite{grandati4}, the potential can be written:

\begin{equation}
V(y,a)=\lambda \left( \lambda \mp \alpha \right) y^{2}+\frac{\mu \left( \mu
-\alpha \right) }{y^{2}}+\lambda _{0}\left( a\right) ,  \label{potcat2+}
\end{equation}%
where $a=(\lambda ,\mu )$ and:

\begin{equation}
\lambda _{0}\left( a\right) =-\alpha \left( \lambda \pm \mu \right)
-2\lambda \mu .
\end{equation}

$V$ is a  translationally shape invariant potential and its energy spectrum
is given by \cite{grandati4}:

\begin{equation}
E_{n}(a)=\pm \left( \phi _{2,\pm }\left( a_{n}\right) -\phi _{2,\pm }\left(
a\right) \right) ,  \label{spectrecat2}
\end{equation}%
where $a=a_{0}$, $a_{k}=(\lambda _{k},\mu _{k})=\left( \lambda \pm k\alpha
,\mu +k\alpha \right) $ and:

\begin{equation}
\phi _{2,\pm }\left( a\right) =\phi _{2,\pm }(\lambda ,\mu )=\left( \lambda
\pm \mu \right) ^{2}.  \label{phi2}
\end{equation}

The classical turning points $y_{i,n}$ for an energy $E_{n}$ are determined
by the condition:

\begin{equation}
E_{n}=V(y_{i,n})\Leftrightarrow \lambda \left( \lambda \mp \alpha \right)
y_{i,n}^{4}+\left( \lambda _{0}\left( a\right) -E_{n}\right) y_{i,n}^{2}+\mu
\left( \mu -\alpha \right) =0,  \label{turning points2}
\end{equation}%
that is:

\begin{equation}
\left\{ 
\begin{array}{c}
y_{2,n}^{2}=u_{0,n}+\delta _{n} \\ 
y_{1,n}^{2}=u_{0,n}-\delta _{n}%
\end{array}%
\right.  \label{centrage1}
\end{equation}%
with:

\begin{equation}
u_{0,n}=\frac{E_{n}-\lambda _{0}\left( a\right) }{2\lambda \left( \lambda
\mp \alpha \right) },\ \delta _{n}=\sqrt{\left( \frac{E_{n}-\lambda
_{0}\left( a\right) }{2\lambda \left( \lambda \mp \alpha \right) }\right)
^{2}-\frac{\mu \left( \mu -\alpha \right) }{\lambda \left( \lambda \mp
\alpha \right) }}.  \label{centrage2}
\end{equation}

In the case$\ dy/dx=\alpha +\alpha y^{2}$, the half action variable for a
classical periodic orbit of energy $E_{n}$ is:

\begin{eqnarray}
\int_{x_{1,n}}^{x_{2,n}}p_{n}(x)dx &=&\int_{y_{1,n}}^{y_{2,n}}\sqrt{%
E_{n}-V(y,a)}\frac{dy}{\alpha \left( y^{2}+1\right) } \\
&=&\frac{\sqrt{\lambda \left( \lambda -\alpha \right) }}{2\alpha }%
\int_{y_{1,n}^{2}}^{y_{2,n}^{2}}duf_{2,+}(u),  \notag
\end{eqnarray}%
where we have defined $u=y^{2}$ and:

\begin{equation}
f_{2,+}(z)=\frac{\sqrt{\left( z-y_{1,n}^{2}\right) \left(
y_{2,n}^{2}-z\right) }}{z\left( z+1\right) }.
\end{equation}

This last integral is readily calculated by using a complex variable
formalism. We have indeed:

\begin{equation}
\int_{y_{1,n}^{2}}^{y_{2,n}^{2}}duf_{2,+}(u)=\pi i\left( \text{Res}\left(
f_{2,+}(z),0\right) +\text{Res}\left( f_{2,+}(z),-1\right) +\text{Res}\left(
f_{2,+}(z),\infty \right) \right) .
\end{equation}

When $z\rightarrow \infty $:

\begin{equation}
f_{2,+}(z)\underset{z\rightarrow \infty }{\sim }-\frac{i}{z}+O(\frac{1}{z^{2}%
}),
\end{equation}%
which implies:

\begin{equation}
\text{Res}\left( f_{2,+}(z),\infty \right) =i.
\end{equation}

As for the residues in $0$ and $-1$, from Eq.(\ref{turning points2}) we have:%
\begin{equation}
\text{Res}\left( f_{2,+}(z),0\right) =\sqrt{-y_{1,n}^{2}y_{2,n}^{2}}=i\sqrt{%
\frac{\mu \left( \mu -\alpha \right) }{\lambda \left( \lambda -\alpha
\right) }}.
\end{equation}

Using Eq.(\ref{turning points2}), Eq.(\ref{centrage1}), Eq.(\ref{centrage2})
and Eq.(\ref{spectrecat2}), we deduce ($-1<0<y_{1,n}^{2}<y_{2,n}^{2}$):

\begin{equation}
\text{Res}\left( f_{2,+}(z),-1\right) =-\sqrt{-\left( 1+y_{1,n}^{2}\right)
\left( y_{2,n}^{2}+1\right) }=\frac{-i\left( \lambda _{n}+\mu _{n}\right) }{%
\sqrt{\lambda \left( \lambda -\alpha \right) }}.
\end{equation}

Then:

\begin{equation}
\int_{x_{1,n}}^{x_{2,n}}p_{n}(x)dx=n\pi +\gamma
\end{equation}%
with:

\begin{equation}
\gamma =\frac{\pi \lambda }{2\alpha }\left( 1-\sqrt{1-\frac{\alpha }{\lambda 
}}\right) +\frac{\pi \mu }{2\alpha }\left( 1-\sqrt{1-\frac{\alpha }{\mu }}%
\right) .  \label{maslov21}
\end{equation}

\bigskip A similar analysis can be led in the case $dy/dx=\alpha -\alpha
y^{2}(x)$. If we except the specific case of the Scarf II potential \cite%
{grandati4}, we can write:

\begin{equation}
\int_{x_{1,n}}^{x_{2,n}}p_{n}(x)dx=-\frac{\sqrt{\lambda \left( \lambda
+\alpha \right) }}{2\alpha }\int_{y_{1,n}^{2}}^{y_{2,n}^{2}}duf_{2,-}(u),
\end{equation}%
where we have defined $u=y^{2}$ and:

\begin{equation}
f_{2,-}(z)=\frac{\sqrt{\left( z-y_{1,n}^{2}\right) \left(
y_{2,n}^{2}-z\right) }}{z\left( z-1\right) }.
\end{equation}

We have:

\begin{equation}
\int_{y_{1,n}^{2}}^{y_{2,n}^{2}}duf_{2,-}(u)=\pi i\left( \text{Res}\left(
f_{2,-}(z),0\right) +\text{Res}\left( f_{2,-}(z),1\right) +\text{Res}\left(
f_{2,-}(z),\infty \right) \right) .
\end{equation}

When $z\rightarrow \infty $:

\begin{equation}
f_{2,-}(z)\underset{z\rightarrow \infty }{\sim }-\frac{i}{z}+O(\frac{1}{z^{2}%
}),
\end{equation}%
which implies:

\begin{equation}
\text{Res}\left( f_{2,-}(z),\infty \right) =i.
\end{equation}

As for the residues in $0$ and $1$, using Eq.(\ref{turning points2}), Eq.(%
\ref{centrage1}), Eq.(\ref{centrage2}) and Eq.(\ref{spectrecat2}), we deduce:%
\begin{equation}
\text{Res}\left( f_{2,-}(z),0\right) =-\sqrt{-y_{1,n}^{2}y_{2,n}^{2}}=-i%
\sqrt{\frac{\mu \left( \mu -\alpha \right) }{\lambda \left( \lambda +\alpha
\right) }}
\end{equation}%
and

\begin{equation}
\text{Res}\left( f_{2,-}(z),1\right) =\sqrt{\left( 1-y_{1,n}^{2}\right)
\left( y_{2,n}^{2}-1\right) }=\frac{-i\left( \lambda _{n}-\mu _{n}\right) }{%
\sqrt{\lambda \left( \lambda +\alpha \right) }}.
\end{equation}

Note that in this case, we had to change the square root determination when
we pass from $0$ to $1$, since we have $0<y_{1,n}^{2}<y_{2,n}^{2}$ $<1$ ( $%
y=\tanh (\alpha x+\varphi )$).

Then:

\begin{equation}
\int_{x_{1,n}}^{x_{2,n}}p_{n}(x)dx=n\pi +\gamma
\end{equation}%
with:

\begin{equation}
\gamma =\frac{\pi \mu }{2\alpha }\left( 1-\sqrt{1-\frac{\alpha }{\mu }}%
\right) -\frac{\pi \lambda }{2\alpha }\left( 1-\sqrt{1+\frac{\alpha }{%
\lambda }}\right) .  \label{maslov22}
\end{equation}

In the case of the Scarf II potential, the variable $y(x)=\tanh \left(
\alpha x/2+i\pi /4\right) $ is a pure phase factor ($\left\vert y\right\vert
=1$) and the path of integration in the action variable surrounds the branch
cut which is an arc of the unit circle. Nevertheless, we can follow the same
reasoning as before and we recover the result given in Eq.(\ref{maslov22}).

Eq.(\ref{maslov21}) and Eq.(\ref{maslov22}) correspond to the results
obtained in \cite{Bhaduri} for the second class potentials.

\section{Conclusion}

We have shown how to calculate exactly and in a general way the action
variable for the whole set of translationally shape invariant potentials.
The correction to lowest order WKB formula appears as a constant term
redefining the Maslov index. This result implies immediately the exactness
of the Ma-Xu formula for every TSIP. The employed techniques of complex
analysis are standard but even so instructive. The\ two basics cases are the
harmonic and isotonic ones (for which we have an equispaced quantum spectrum
and isochronicity at the classical level).\ In this two cases, the action
integral involved is a sum of at most two residues and the linear $n$
dependence ($n$ being the energy quantum number) comes from the residue at
infinity. For all the other cases, the integral involved\ is obtained from
the two basic cases by deforming the integration measure with a weight which
is the inverse of an at most quadratic polynomial. The residue at infinity
becomes energy independent and the linear $n$ dependence takes its origin in
the finite poles (with an eventual compensation of nonlinear $n$ dependent
contributions).

\section{Acknoledgments}

We would like to thank Y. Kasri for calling our attention to the Ma-Xu
formula and for stimulating and useful discussions.



\begin{thebibliography}{99}
\bibitem{Maxu1} Z.Q. Ma and B.\ W.\ Xu, \textquotedblleft Exact quantization
rules for bound states of the Schr\"{o}dinger equation\textquotedblright ,\
Int. J. Mod. Phys. E, \textbf{14}, 599-610 (2005).

\bibitem{Maxu2} Z.Q. Ma and B.\ W.\ Xu, \textquotedblleft Quantum correction
in exact quantization rules\textquotedblright ,\ Europhys. Lett. \textbf{69}%
, 685-691 (2005)

\bibitem{Cao} Z.\ Q.\ Cao, Q. Liu, Q.\ S.\ Shen, X.\ Dou, Y.\ Chen and Y.\
Ozaki, \textquotedblleft Quantization scheme for arbitrary one-dimensional
potential wells\textquotedblright ,\ \ Phys. Rev. A \textbf{63}, 054103
(2001)

\bibitem{Cao2} F. Zhou, Z.\ Q.\ Cao and Q.\ S.\ Shen, \textquotedblleft
Energy splitting in symmetric double-well potentials\textquotedblright ,\
Phys. Rev. A \textbf{67}, 062112 (2003)

\bibitem{Ma} Z.Q. Ma, A. Gonzalez-Cisneros, B.\ W.\ Xu and S. H. Dong,
\textquotedblleft Energy spectrum of the trigonometric Rosen--Morse
potential using an improved quantization rule\textquotedblright ,\ Phys.
Lett. A \textbf{371}, 180-184 (2007).

\bibitem{qiang} W. C. Qiang and S. H. Dong, \textquotedblleft Arbitrary
l-state solutions of the rotating Morse potential through the exact
quantization rule method\textquotedblright ,\ Phys. Lett. A \textbf{363},
169-176 (2007).

\bibitem{qiang2} W. C. Qiang, R. S. Zhou and Y. Gao , \textquotedblleft
Application of the exact quantization rule to the relativistic solution of
the rotational Morse potential with pseudospin symmetry\textquotedblright ,\
J. Phys. A, \textbf{40}, 1677-1685 (2007).

\bibitem{qiang3} W. C. Qiang, Y. Gao and R. S. Zhou , \textquotedblleft\
Arbitrary l-state approximate solutions of the Hulth\'{e}n through potential
the exact quantization rule\textquotedblright ,\ Cent. Eur. J. Phys., 
\textbf{6}, 356-362 (2008).

\bibitem{gu} X. Y. Gu and S. H. Dong, \textquotedblleft The improved
quantization rule and the Langer modification\textquotedblright ,\ Phys.
Lett. A \textbf{372}, 1972-1977 (2008).

\bibitem{gu2} X. Y. Gu, S. H. Dong and Z. Q. Ma , \textquotedblleft Energy
spectra for modified Rosen--Morse potential solved by the exact quantization
rule\textquotedblright ,\ J. Phys. A, \textbf{42}, 035303 (2009).

\bibitem{dong} S. H. Dong and A. Gonzalez-Cisneros , \textquotedblleft
Energy spectra of the hyperbolic and second P\"{o}schl--Teller like
potentials solved by new exact quantization rule\textquotedblright ,\ Ann.
Phys., \textbf{323, }1136-1149 (2008).

\bibitem{dong2} S.H. Dong, \textquotedblleft A new quantization rule to the
energy spectra for modified hyperbolic-type potentials\textquotedblright ,\
Int. J. Quant. Chem., \textbf{109}, 701-707 (2009).

\bibitem{dong3} S. H. Dong, D. Morales and J. Garcia-Ravelo,
\textquotedblleft Exact quantization rule and its applications to physical
potentials\textquotedblright ,\ Int. J. Mod. Phys. E, \textbf{16}, 189-198
(2007).

\bibitem{kasri} Y.\ Kasri and L.\ Chetouani, \textquotedblleft\ Application
of the exact quantization rule for some noncentral separable
potentials\textquotedblright ,\ Can. J. Phys., \textbf{86}, 1-7 (2008).

\bibitem{cooper} F.\ Cooper, A.\ Khare and U.\ Sukhatme, \textit{%
Supersymmetry in quantum mechanics} (World Scientific, Singapore, 2001).

\bibitem{comtet} A.\ Comtet, A.D.\ Bandrauk abd D.K.\ Campbell,
\textquotedblleft Exactness of semiclassical bound state energies for
supersymmetric quantum mechanics\textquotedblright ,\ Phys. Lett. B \textbf{%
150}, 159-162 (1985).

\bibitem{dutt2} R.\ Dutt, A. Khare and U.\ P.\ Sukhatme, \textquotedblleft
Supersymmetry-inspired WKB approximation in quantum
mechanics\textquotedblright ,\ Am. J. Phys. 5\textbf{9}, 723--727 (1991).

\bibitem{barclay1} D. T. Barclay and C.J. Maxwell, \textquotedblleft Shape
invariance and the SWKB series\textquotedblright ,\ Phys. Lett. A \textbf{157%
}, 357-360 (1991).

\bibitem{barclay2} D. T. Barclay, \textquotedblleft Convergent WKB
series\textquotedblright ,\ Phys Lett. A \textbf{185}, 169-173 (1994).

\bibitem{Bhaduri} R.K. Bhaduri, J.\ Sakhr, D.W.L.\ Sprung, R.\ Dutt and A.\
Suzuki, \textquotedblleft Shape invariant potentials in SUSY quantum
mechanics and periodic orbit theory\textquotedblright ,\ J. Phys. A, \textbf{%
38}, L183 (2005).

\bibitem{Bhaduri2} R.K. Bhaduri, \textit{Semiclassical physics}
(Westview-Frontiers in Physics, Boulder, 2003).

\bibitem{grandati4} Y. Grandati and A.\ B\'{e}rard, \textquotedblleft
Rational solutions for the Riccati-Schr\"{o}dinger equations associated to
translationally shape invariant potentials,\textquotedblright\
arXiv:0910.4810 .

\bibitem{Dutt} R.\ Dutt, A. Khare and U.\ P.\ Sukhatme, \textquotedblleft
Supersymmetry, shape invariance and exactly solvable
potentials\textquotedblright ,\ Am. J. Phys. 5\textbf{6}, 163--168 (1988).

\bibitem{bhalla} R.\ S.\ Bhalla, A.\ K.\ Kapoor and P.\ K.\ Panigrahi,
\textquotedblleft Exactness of the supersymmetric WKB approximation
scheme\textquotedblright ,\ Phys. Rev. A, \textbf{54}, 951-954 (1996).

\bibitem{bhalla2} R.\ S.\ Bhalla, A.\ K.\ Kapoor and P.\ K.\ Panigrahi,
\textquotedblleft Energy Eigenvalues for a Class of One-Dimensional
Potentials via Quantum Hamilton-Jacobi Formalism\textquotedblright ,\ Mod.
Phys. Lett. A, \textbf{12}, 295-306 (1997).

\bibitem{bhalla3} R.\ S.\ Bhalla, A.\ K.\ Kapoor and P.\ K.\ Panigrahi,
\textquotedblleft Quantum Hamilton--Jacobi formalism and the bound state
spectra\textquotedblright ,\ Am. J. Phys., \textbf{65}, 1187-1194 (1997).

\bibitem{cherqui} C.\ Cherqui, Y.\ Binder and A.\ Gangopadhyaya,
\textquotedblleft Shape invariance and the exactness of the Quantum
Hamilton--Jacobi formalism\textquotedblright , Phys. Lett. A \textbf{372}
1406-1415 (2008).

\bibitem{leacock1} R.\ A.\ Leacock and M.\ J.\ Padgett, \textquotedblleft
Hamilton-Jacobi Theory and the Quantum Action Variable\textquotedblright ,\
Phys. Rev. Lett. \textbf{50}, 3 - 6 (1983).

\bibitem{leacock2} R.\ A.\ Leacock and M.\ J.\ Padgett, \textquotedblleft
Hamilton-Jacobi/action-angle quantum mechanics\textquotedblright ,\ Phys.
Rev. D \textbf{28}, 2491 - 2502 (1983).

\bibitem{weissman} Y. Weissman and J. Jortner, \textquotedblleft The
isotonic oscillator\textquotedblright , Phys. Lett. A 70, 177--179 (1979).
\end{thebibliography}
\end{document}